\documentclass[11pt,letterpaper]{article}
\usepackage{systeme} %para usar sistemas de ecuaciones
\usepackage[utf8]{inputenc}
\usepackage{multirow} %para las tablas
\usepackage{booktabs} %para la estética de tablas
\usepackage[english]{babel}
\usepackage{amsmath}
\usepackage{amsfonts}
\usepackage{amssymb}
\usepackage{graphicx}
\usepackage[small,bf]{caption} %Para caption con letra pequeña y Figura en negrita
\usepackage[left=3cm,right=3cm,top=2cm,bottom=3cm]{geometry}

\author{Jorge Pinochet}
\title{\textbf{General relativity in a nutshell I}}
\begin{document}

\author{Jorge Pinochet$^{*}$\\ \\
 \small{$^{*}$\textit{Facultad de Ciencias Básicas, Departamento de Física. }}\\
  \small{\textit{Centro de Investigación en Educación (CIE-UMCE),}}\\
 \small{\textit{Núcleo Pensamiento Computacional y Educación para el Desarrollo Sostenible (NuCES).}}\\
 \small{\textit{Universidad Metropolitana de Ciencias de la Educación,}}\\
 \small{\textit{Av. José Pedro Alessandri 774, Ñuñoa, Santiago, Chile.}}\\
 \small{e-mail: jorge.pinochet@umce.cl}\\}

\date{}
\maketitle

\begin{center}\rule{0.9\textwidth}{0.1mm} \end{center}
\begin{abstract}
\noindent Einstein's general relativity is the best available theory of gravity. In recent years, spectacular proofs of Einstein's theory have been conducted, which have aroused interest that goes far beyond the narrow circle of specialists. The aim of this work is to offer an elementary introduction to general relativity. In this first part, we introduce the geometric concepts that constitute the basis of Einstein's theory. In the second part we will use these concepts to explore the curved spacetime geometry of general relativity.\\ \\


\begin{center}\rule{0.9\textwidth}{0.1mm} \end{center}
\end{abstract}

\maketitle

\section{Introduction}

General relativity (GR) is the best available theory of gravity. Although GR aroused great interest in its early years, especially after the results of the 1919 British astronomical expeditions corroborating Einstein's ideas became known, interest quickly waned and for the next five decades physicists directed their attention to other issues. In the 1960s, the situation changed dramatically, and the theory of GR underwent a renaissance. This is the time that the theoretical physicist Kip Thorne calls the golden age of GR\footnote{During this time, new mathematical techniques were developed that made calculations easier, technology had advanced enough to allow accurate tests of GR to be carried out, and astronomical observations were beginning to reveal extreme phenomena that could only be adequately explained in the framework of GR (such as pulsars and quasars), so that physicists and astronomers began to pay attention to Einstein’s theory.} [1]. In recent years, GR has experienced an extraordinary boom during which spectacular proofs of Einstein's theory have been presented that have aroused interest that goes beyond the narrow circle of specialists. Three milestones in this regard include the first detection of gravitational waves made by the Laser Interferometer Gravitational-Wave Observatory (LIGO) collaboration in 2015, the first image of a black hole obtained by the Event Horizon Telescope (EHT) collaboration in 2019, and the spectacular image obtained in 2022 by the same collaboration of the black hole that inhabits in the centre of our galaxy.\\

In this scenario, it does not seem risky to affirm that GR is experiencing a second golden age [2], which provides a great opportunity for specialists and educators in the area to try to explain the meaning and scope of GR to a public as wide as possible. However, this great opportunity also represents a great challenge, since GR has a well-deserved reputation for being a highly mathematically complex theory. The objective of this work is thus to is to meet this challenge by offering an elementary introduction to GR.\\

The article has been divided into two parts and is aimed at those who have mastered infinitesimal calculus, as well as the fundamentals of the theory of special relativity and Newtonian gravitation. The objective of this first part is to introduce the mathematical and geometric concepts that constitute the basis of GR, which is a geometric theory of gravity that describes the interaction between massive bodies as an effect of the curvature of spacetime. In the second part, the ideas developed here are used to explore this curved spacetime geometry.

\section{Riemann geometry and the concept of metric}

Euclidean geometry is the branch of mathematics that studies the properties of flat spaces. Curved spaces are described by non-Euclidean geometry. As will later be shown in detail, the essential difference is that only Euclidean geometry satisfies the parallel postulate, which states that two initially parallel lines that are prolonged indefinitely maintain their mutual distance constant.\\

The study of non-Euclidean geometries on curved surfaces began in the 19th century with the work of Lobachevski, Bolyai, and Gauss. These works were generalised by Riemann for the study of curved spaces in any dimension [3]. The fundamental research object of Riemann's geometry is the \textit{manifold}, which generalizes the notions of curve (one-dimensional manifold) and surfaces (two-dimensional manifold) to spaces having more than two dimensions ($n$-dimensional manifolds). Riemannian geometry includes Euclidean geometry as a particular case of zero curvature.

\begin{figure}[h]
  \centering
    \includegraphics[width=0.4\textwidth]{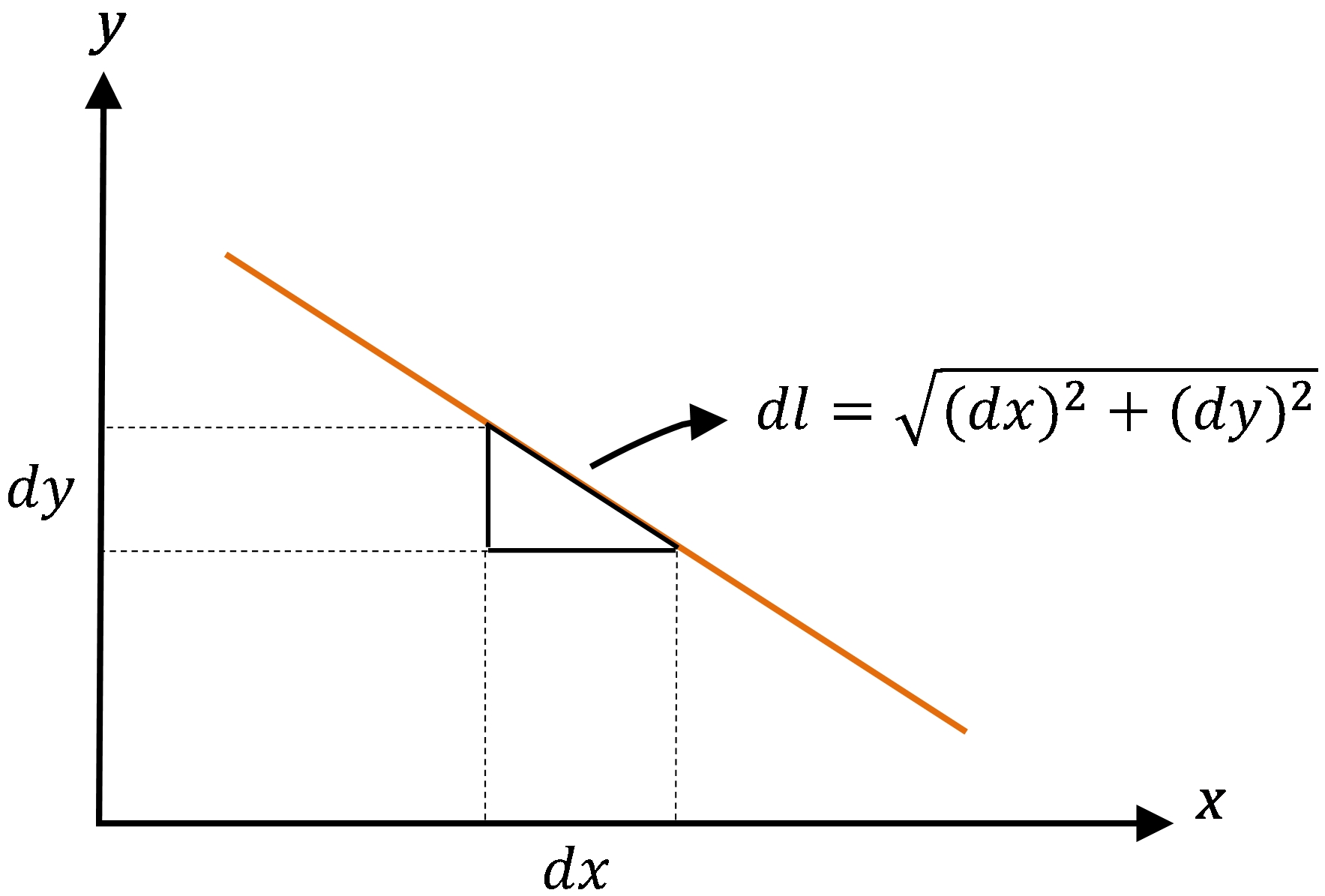}
  \caption{The Pythagorean theorem in differential form.}
\end{figure}

Riemannian geometry is the mathematical foundation of GR. One of the fundamental principles of this geometry is that \textit{in a small neighbourhood of a point, non-Euclidean manifolds agree with Euclidean geometry}. In other words, \textit{a sufficiently small region of curved space can be considered flat to a first approximation}. An important corollary to this principle is that the Pythagorean theorem of Euclidean geometry holds locally on a non-Euclidean manifold. The fundamental mathematical object of Riemannian geometry is the \textit{metric}, which is a powerful generalisation of the Pythagorean theorem. The metric allows the calculation of the infinitesimal distance or line element $dl$ between two points on any \textit{n}-dimensional manifold [3]:

\begin{equation} %1
(dl)^{2} = \sum_{i=1}^{n} \sum_{j=1}^{n} g_{ij} dx_{i} dx_{j}.
\end{equation}

where $dx_{1}, dx_{2},... dx_{n}$ are the differentials of the coordinates $x_{1}, x_{2},... x_{n}$, and $g_{ij}$ are a set of quantities known as the \textit{metric tensor}, which are functions of the coordinates and contain all the information about the manifold. An important aspect of Eq. (1) is that $dl$ is an invariant, that is, the infinitesimal distance between two points cannot depend on the chosen coordinate system. The set of $g_{ij}$ is represented by an $n\times n$ square matrix that has $n^{2}$ components [3,4]:

\begin{equation} %2
\left[g_{ij} \right]  = 
\begin{pmatrix}
g_{11} & g_{12} & \cdots & g_{1n} \\
g_{21} & g_{22} & \cdots & g_{2n} \\
\vdots  & \vdots  & \ddots & \vdots  \\
g_{n1} & g_{n2} & \cdots & g_{nn} 
\end{pmatrix}.
\end{equation}

This matrix is symmetric, that is, the components under the main diagonal are equal to the components on said diagonal ($g_{ij}=g_{ji}$). This implies that the number $N$ of independent components is less than $n^{2}$, and is calculated as:

\begin{equation}%3
N = 1+2+... +n = \frac{1}{2} n(n+1).
\end{equation}

To understand the meaning of Eqs. (1) and (2), let us consider the simple case of a two-dimensional manifold ($n = 2$). Under this condition, the metric tensor is:

\begin{equation} %4
\left[g_{ij} \right] = 
\begin{pmatrix}
g_{11} & g_{12}\\
g_{21} & g_{22} 
\end{pmatrix},
\end{equation}

where the metric is written as:

\begin{equation} %5
(dl)^{2} = \sum_{i=1}^{2} \sum_{j=1}^{2} g_{ij} dx_{i} dx_{j} = g_{11}dx_{1}dx_{1} + g_{22}dx_{2}dx_{2} + g_{12}dx_{1}dx_{2} + g_{21}dx_{2}dx_{1}.
\end{equation}

Suppose that the manifold described by this expression has zero curvature (is a plane). If Cartesian coordinates $x,y$ are used then $x_{1} = x, x_{2}=y;\ g_{11}=g_{22}=1, g_{12}=g_{21}=0$, so,

\begin{equation} %6
\left[g_{ij} \right] = 
\begin{pmatrix}
1 & 0\\
0 & 1 
\end{pmatrix},
\end{equation}

and the metric is the Pythagorean theorem in two dimensions (Fig. 1):

\begin{equation} %7
(dl)^{2} = (1)dxdx + (1)dydy + (0)dxdy + (0)dydx = (dx)^{2} + (dy)^{2}.
\end{equation}

Adopting the convention of removing parentheses this equation can be rewritten as:

\begin{equation} %8
dl^{2} = dx^{2} + dy^{2}.
\end{equation}

Another simple example is the metric described by this equation but expressed in polar coordinates $r, \varphi$ (see Fig. 2). In this case, $x_{1} = r, x_{2} = \varphi;\ g_{11} = 1, g_{22}=r^{2}, g_{12}=g_{21}=0$, where:

\begin{equation} %9
\left[g_{ij} \right] = 
\begin{pmatrix}
1 & 0\\
0 & r^{2} 
\end{pmatrix},
\end{equation}

and applying the same logic that led us to Eq. (8) we obtain:

\begin{equation}%10
dl^{2} = dr^{2} + r^{2}d\varphi^{2}.
\end{equation}

\begin{figure}[h]
  \centering
    \includegraphics[width=0.35\textwidth]{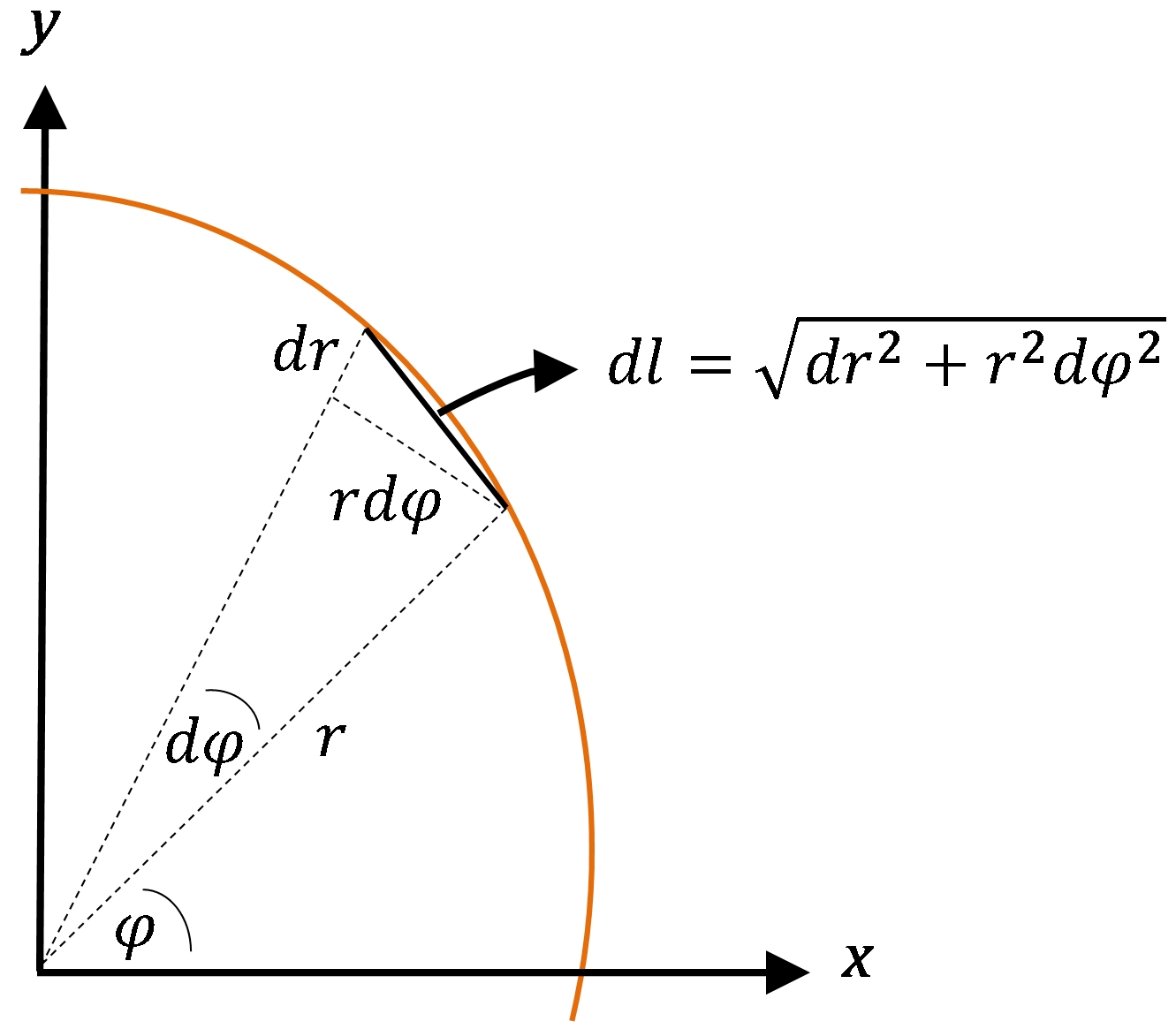}
  \caption{Geometric method to obtain the metric of a plane in polar coordinates.}
\end{figure}

Let us now analyse the simplest case of a curved manifold, the two-dimensional surface of a sphere ($n = 2$), which can be described by the spherical coordinates $\theta, \varphi$ where $0\leq \theta \leq \pi$ is the polar angle, $0\leq \varphi \leq 2\pi$ is the equatorial angle, and the radius $r > 0$ is constant (Fig. 3). Taking $x_{1} = \theta, x_{2} = \varphi;\ g_{11} = r^{2}, g_{22}=r^{2}\sin^{2}\theta, g_{12}=g_{21}=0$ results in:

\begin{equation} %11
\left[g_{ij} \right] = 
\begin{pmatrix}
r^{2} & 0\\
0 & r^{2}\sin^{2}\theta 
\end{pmatrix},
\end{equation}

then,

\begin{equation}%12
dl^{2} = r^{2}d\theta^{2} + r^{2}\sin^{2}\theta d\varphi^{2}.
\end{equation}

Although the surface of the sphere is curved, this metric was obtained using the Pythagorean theorem of plane geometry (see Fig. 3), considering that non-Euclidean manifolds agree with Euclidean geometry in a small neighbourhood of a point.\\

Consider the case of a three-dimensional Euclidean manifold ($n = 3$). In Cartesian coordinates $x,y,z$ we have $x_{1} = x, x_{2} = y, x_{3} = z;\ g_{11} =  g_{22}= g_{33} = 1, g_{ij} = 0$ if $i \neq j$, where,

\begin{equation} %13
\left[g_{ij} \right] = 
\begin{pmatrix}
1 & 0 & 0\\
0 & 1 & 0\\
0 & 0 & 1
\end{pmatrix}.
\end{equation}

and the metric is the Pythagorean theorem in three dimensions:

\begin{equation}%14
dl^{2} = dx^{2}+ dy^{2} + dz^{2}.
\end{equation}

This equality can also be expressed in spherical coordinates $r,\theta,\varphi$; however, unlike Eq. (11), now the radius $r$ is variable. Under these conditions $x_{1} = r, x_{2} = \theta, x_{3} = \varphi;\ g_{11} = 1, g_{22}=r^{2} ,g_{33} = r^{2}\sin^{2}\theta, g_{ij} = 0$ if $i \neq j$, where (see Fig. 3),

\begin{equation} %15
\left[g_{ij} \right] = 
\begin{pmatrix}
1 & 0 & 0\\
0 & r^{2} & 0\\
0 & 0 & r^{2}\sin^{2}\theta
\end{pmatrix}.
\end{equation}

Therefore,

\begin{equation}%16
dl^{2} = dr^{2} + r^{2}d\theta^{2} + r^{2} \sin^{2}\theta d\varphi^{2}.
\end{equation}

If $r = constant$ then $dr = 0$, and, as expected, the spherical surface metric defined by Eq. (12) is recovered.

\begin{figure}[h]
  \centering
    \includegraphics[width=0.5\textwidth]{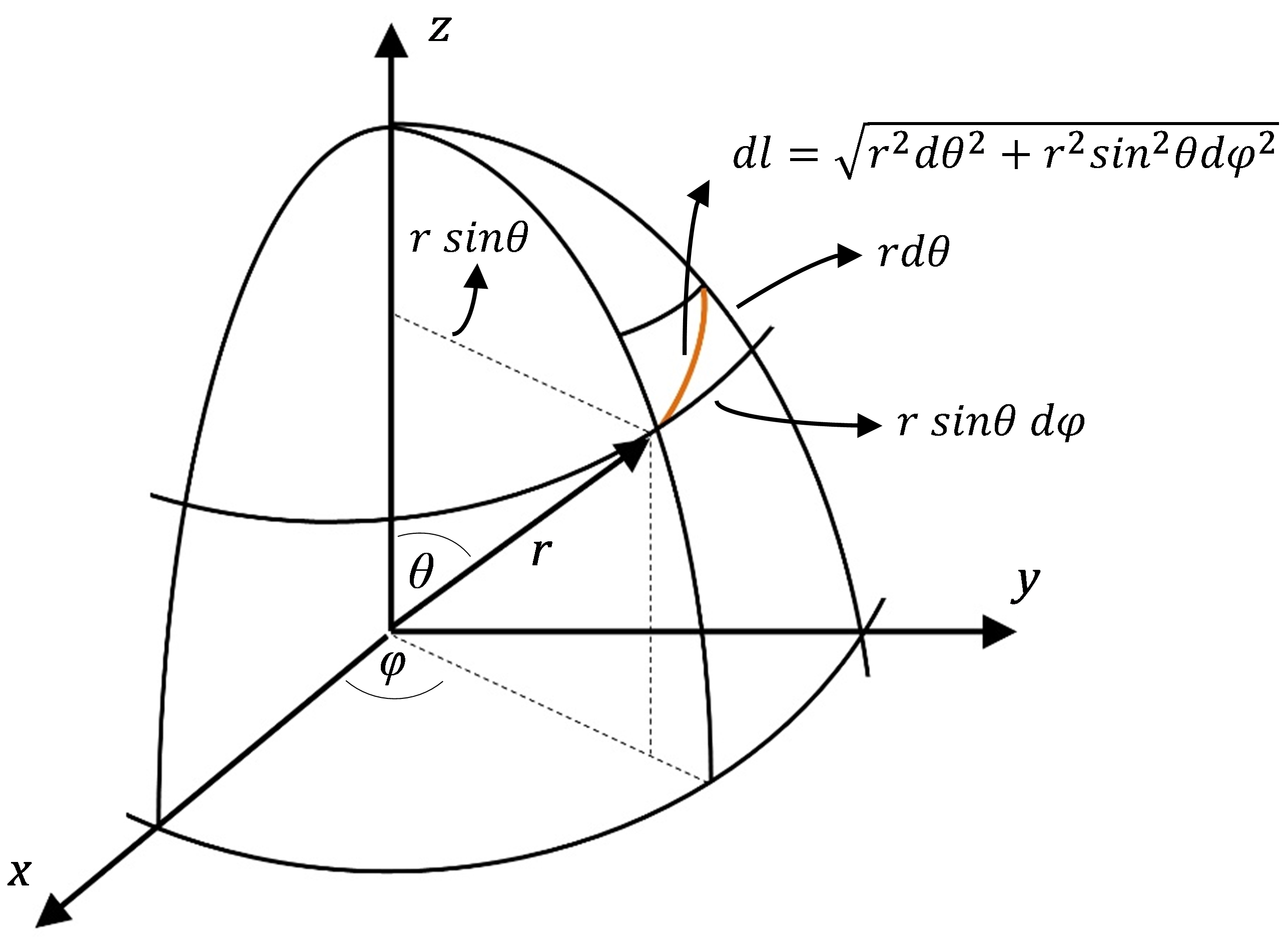}
  \caption{Geometric method to obtain the metric of the surface of a sphere. If $r$ is variable, this figure allows the metric of a three-dimensional flat space to be obtained, Eq. (16).}
\end{figure}

All the properties of a manifold can be calculated from the metric or the metric tensor. This tensor is also used to perform coordinate transformations, where a line element $dl$ written in a system $x_{1}, x_{2},... x_{n}$ can be transformed into a line element $dl'$ written in a system $x_{1}', x_{2}',...x_{n}'$ under the condition $dl = dl'$. For example, since Eq. (8) describes a two-dimensional Euclidean manifold in $x,y$ coordinates, and Eq. (10) describes the same manifold in $r,\varphi$ coordinates, a transformation leading from Eq. (8) to (10) and vice versa can always be performed.\\

When the curvature is zero at every point of the manifold, it is always possible to find a coordinate system where $g_{ij} = 1$ if $i=j$, and $g_{ij}=0$ if $i \neq j$, so that the metric tensor takes the simple diagonal form:

\begin{equation} %17
\left[g_{ij} \right]  = 
\begin{pmatrix}
1 & 0 & \cdots & 0 \\
0 & 1 & \cdots & 0 \\
\vdots  & \vdots  & \ddots & \vdots  \\
0 & 0 & \cdots & 1 
\end{pmatrix}.
\end{equation}

Two particular cases of this expression are Eqs. (6) and (13).\\

Finally, it is worth bearing in mind that, although for simplicity we have only considered examples of diagonal metric tensors, where there are only non-zero components on the main diagonal of the matrix, there are also non-diagonal tensors which are generally associated with coordinate systems whose axes are not mutually orthogonal.

\section{Curvature and geodesics: An intuitive look}

The geometric properties of a manifold are invariant under coordinate transformations. \textit{Curvature} and \textit{geodesics} are two central properties of a manifold that are invariant. Like other geometric properties of a manifold, curvature and geodesics can be computed from the metric. Since these calculations are usually quite complex, we will now discuss them intuitively, focusing on two-dimensional manifolds of constant curvature, such as the surface of a sheet of paper or a sphere. However, everything we say about these surfaces is valid for the higher dimensional manifolds.\\

\begin{figure}[h]
  \centering
    \includegraphics[width=0.35\textwidth]{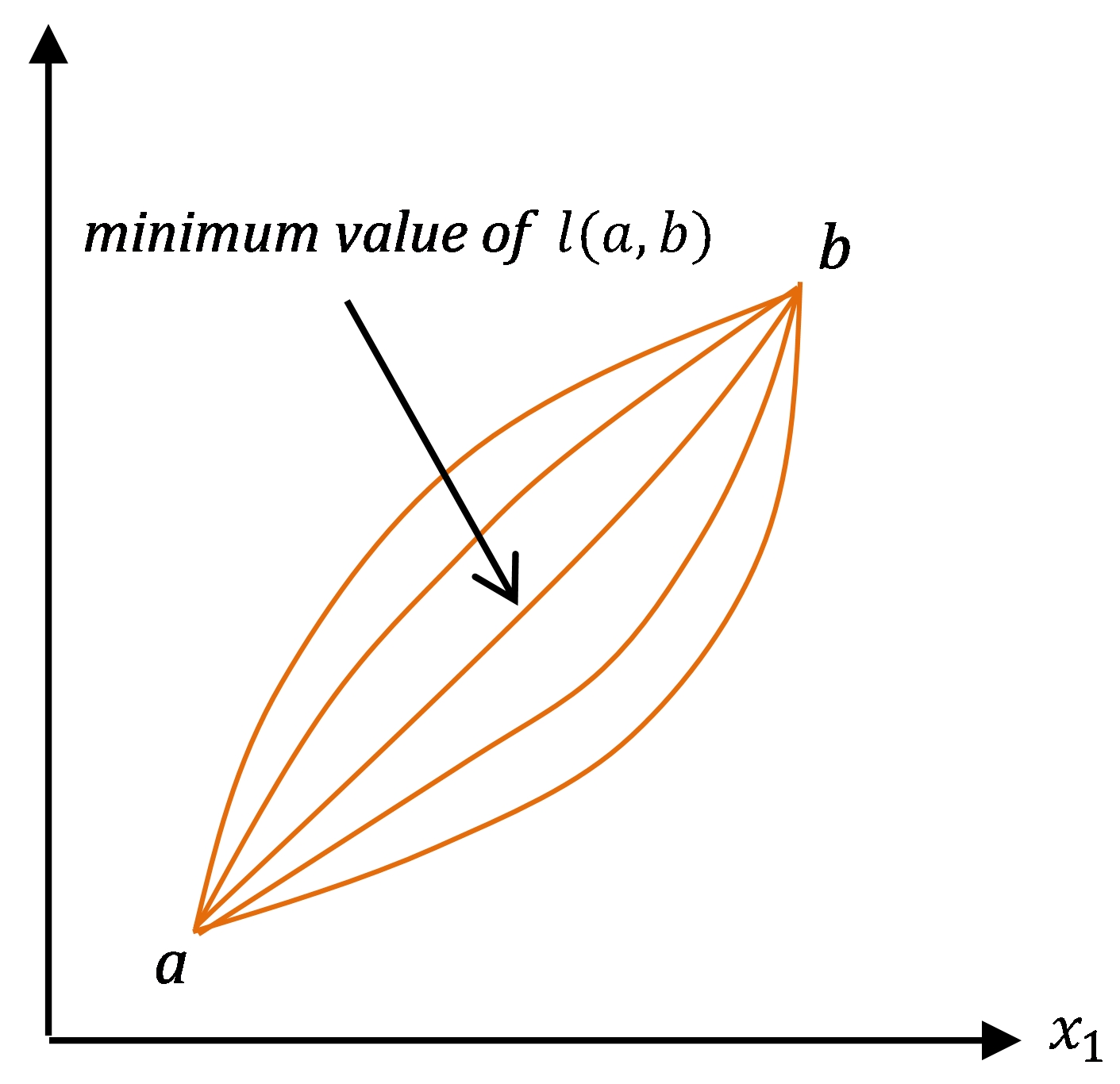}
  \caption{The geodesic $l(a,b)$ is the shortest distance between two points $a$ and $b$.}
\end{figure}

Let us start with the notion of a geodesic, which in Riemann geometry is defined as the locally shortest path between two points on a given manifold, and which is contained in that manifold. An alternative definition of geodesic that will be more useful for this study of GR is that it is a locally straight line. On a flat manifold, the geodesics are straight lines, but on a curved manifold like the surface of a sphere, the geodesics will be a locally straight line. Then, the geodesic generalises the notion of a straight line from Euclidean geometry. The procedure for calculating geodesics consists of taking two fixed points $a$ and $b$ on a given manifold, and then determining which of all the paths connecting $a$ and $b$ is the straightest locally (Fig. 4). According to Eq. (8), in the case of a two-dimensional Euclidean manifold the distance between $a$ and $b$ is:

\begin{equation}%18
\int_{a}^{b} dl = l(a,b) = \int_{a}^{b} \sqrt{dx^{2} + dy^{2}}=\int_{a}^{b}\sqrt{\left(\dfrac{dx}{dt} \right)^{2} +\left( \frac{dx}{dt}\right)^{2}}dt.
\end{equation}

where $t$ is a parameter. Eq. (1) can be used to generalise this expression as [3,4]:

\begin{equation}%19
\int_{a}^{b} dl = l(a,b) = \int_{a}^{b} \sqrt{\sum_{i=1}^{n} \sum_{j=1}^{n} g_{ij} dx_{i} dx_{j}}=\int_{a}^{b}\sqrt{\sum_{i=1}^{n} \sum_{j=1}^{n} g_{ij}\frac{dx_{i}}{dt}\frac{dx_{j}}{dt}}dt,
\end{equation}

A simple example is the surface of a sphere, whose geodesics are the great circles obtained as the intersection of the surface with a plane passing through its centre. In particular, the equator and the meridians are geodesic lines.

\begin{figure}[h]
  \centering
    \includegraphics[width=0.6\textwidth]{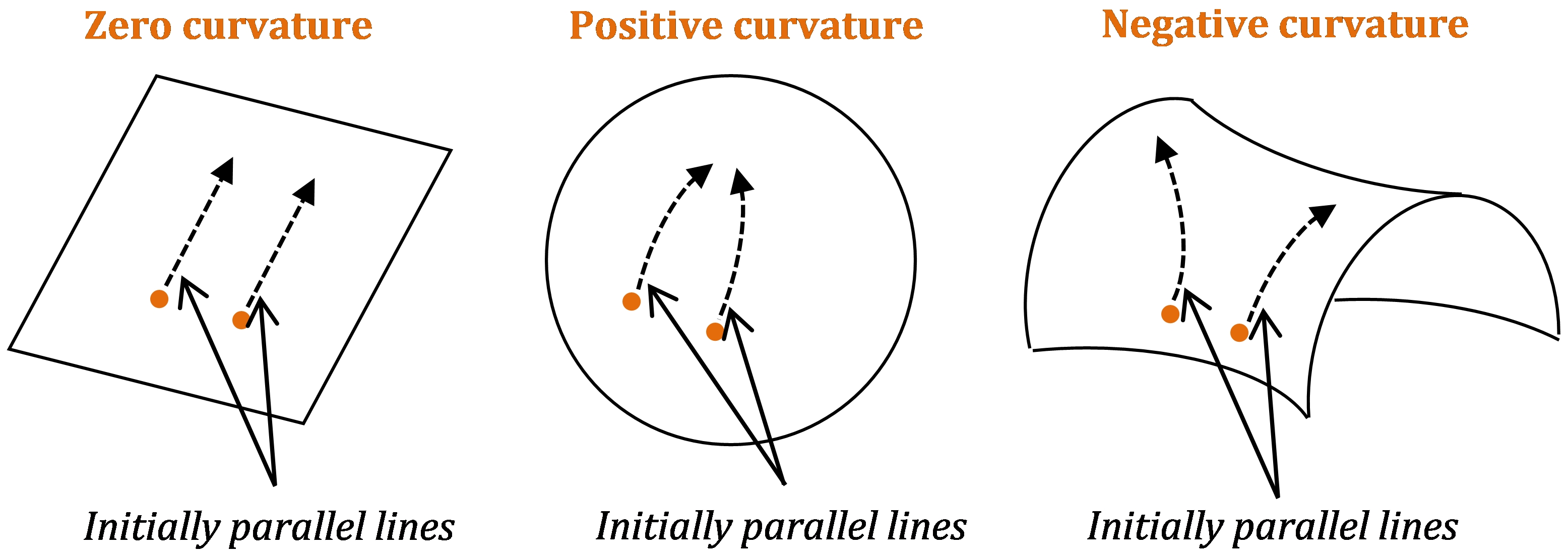}
  \caption{Examples of null (paper sheet), positive (sphere), and negative (saddle) curvatures.}
\end{figure}

Let us now analyse the concept of curvature. As shown in Figure 5, a manifold can have only three curvature classes: null, positive, and negative [3]. Although the surfaces shown in Figure 5 have a constant curvature, it is important to keep in mind that, in most situations, the curvature varies from point to point. However, Figure 5 illustrates the central ideas that we are interested in analysing. Thus, in a flat manifold (Fig. 5 left), two initially parallel geodesics which are indefinitely prolonged keep their mutual distance constant, that is, the parallel postulate is fulfilled. Instead, on a manifold with a positive curvature, two initially parallel geodesics converge (Fig. 5 centre), and on a manifold with a negative curvature, the geodesics diverge (Fig. 5 right).\\

\begin{figure}[h]
  \centering
    \includegraphics[width=0.3\textwidth]{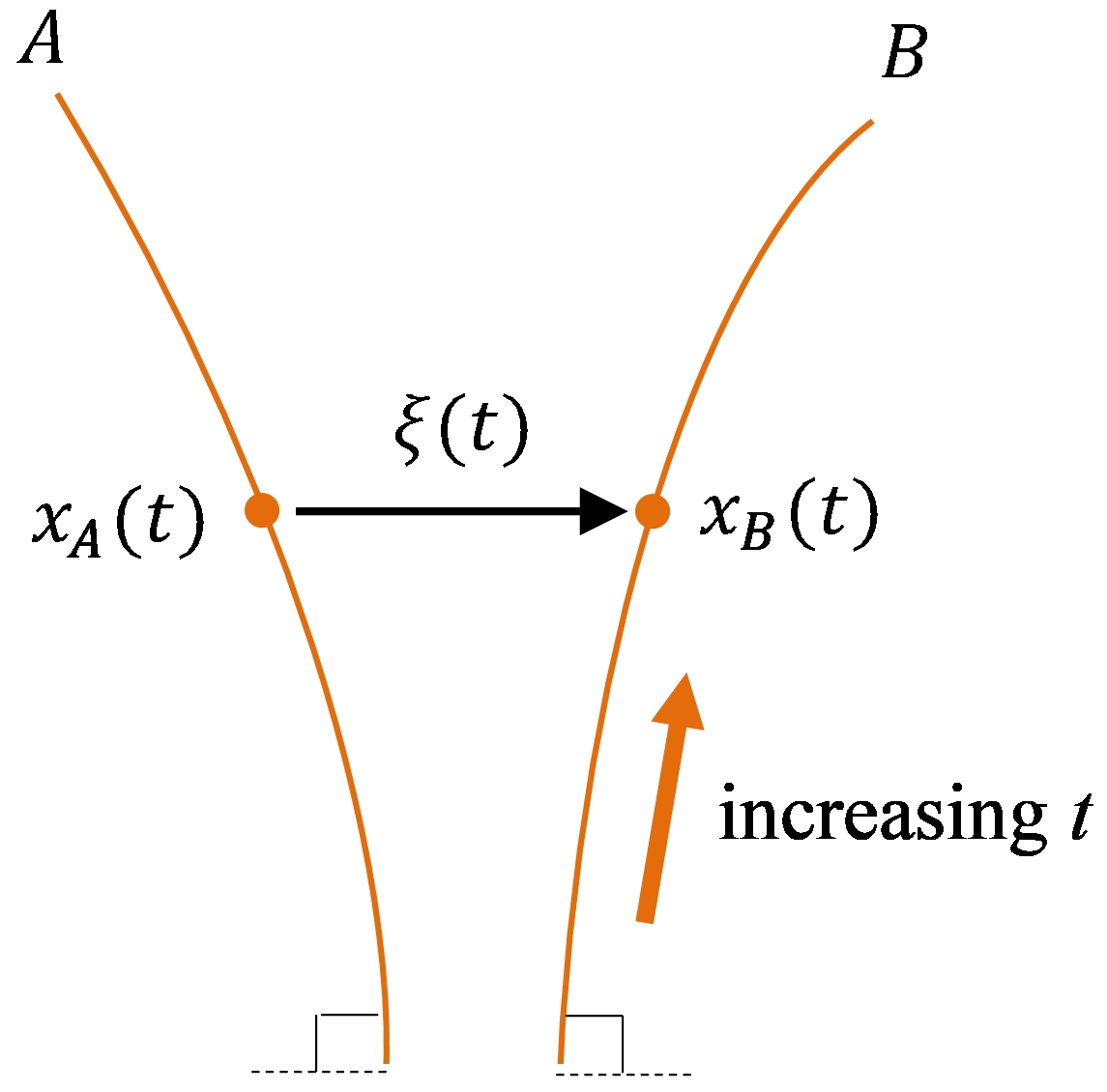}
  \caption{Geodesic deviation $\xi(t)$ between two points $x_{A}(t)$ (geodesic \textit{A}) and $x_{B}(t)$ (geodesic \textit{B}).}
\end{figure}

This reveals a very important result: a single geodesic does not allow the type of curvature of a manifold to be established. To do this, two geodesics are required; in particular, it must be determined if the geodesics converge or diverge, that is, if there is a \textit{geodesic deviation}, which is a measure of the curvature of a manifold. The more pronounced the deviation, the greater the curvature.\\

In Euclidean geometry the geodesic deviation is zero. However, locally, in a non-Euclidean manifold the geodesic deviation is also zero, since, in a small neighbourhood of a point, non-Euclidean manifolds agree with Euclidean geometry. Figure 6 illustrates these ideas, where two geodesics $A$ and $B$ are separated by a distance $\xi(t)$, where $t$ is a parameter. Intuitively, if $x_{A}(t)$ and $x_{B}(t)$ are the positions of the particles in a certain coordinate system, the geodesic deviation is defined as [3,4]:

\begin{equation}%20
\xi(t) = x_{A}(t) - x_{B}(t).
\end{equation}

Then, if $\xi(t)$ decreases over time, we are in the situation illustrated in Figure 5 (centre) where the curvature is positive. If $\xi(t)$ increases, the situation corresponds to figure 5 (right) where the curvature is negative. In these two situations the manifold is non-Euclidean. However, if $\xi(t)$ remains constant over time, the situation corresponds to figure 5 (left) where the curvature is zero and the manifold is Euclidean.

\section{The metric of a non-Euclidean manifold: A simple example}

The basic criterion for establishing whether a manifold is non-Euclidean is to find disagreements with some results of Euclidean geometry, such as those shown in Figure 7. Let us consider, for example, the ratio between the perimeter $C$ and the radius $r$ of a circle (Fig. 7, centre). We can calculate this ratio from Eq. (10), which gives the metric of a two-dimensional Euclidean manifold. To carry out the calculation, a region of constant radius is considered, which implies that $dr = 0$. Integrating between $0$ and $2\pi$:

\begin{equation}%21
\int_{0}^{C} dl = \int_{0}^{2\pi} r d\varphi \rightarrow C=2\pi r.
\end{equation}

\begin{figure}[h]
  \centering
    \includegraphics[width=0.8\textwidth]{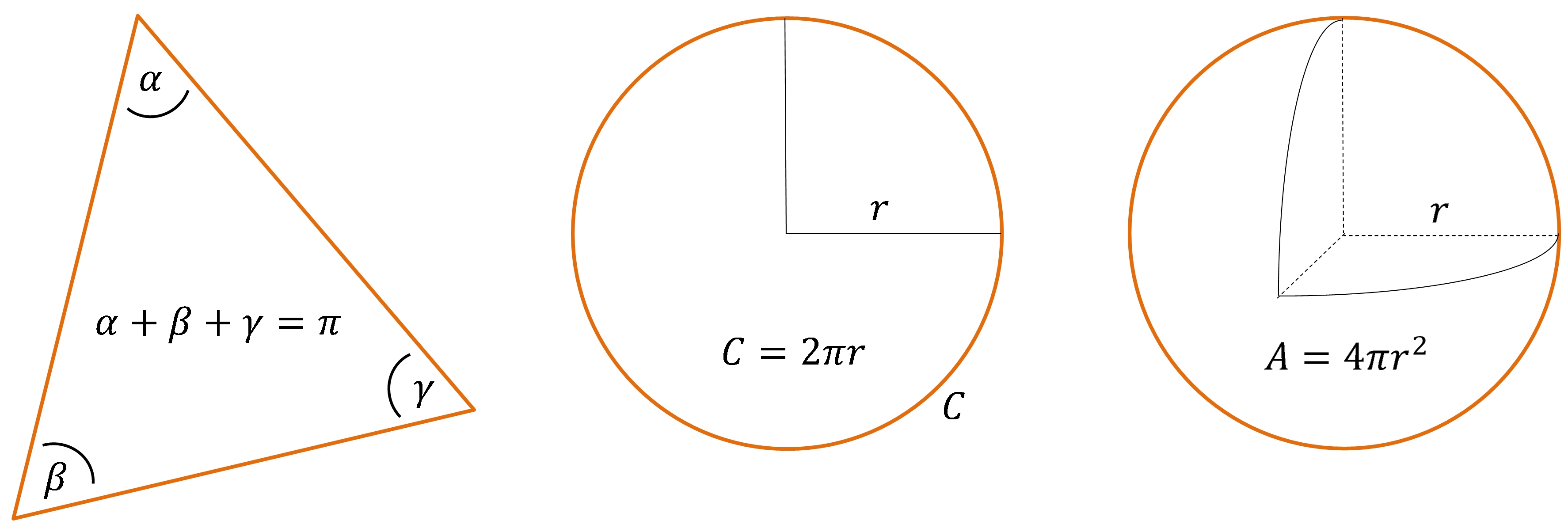}
  \caption{Some well-known results of Euclidean geometry. Left: Sum of the interior angles of a triangle. Center: Perimeter of a circle. Right: Surface area of a sphere.}
\end{figure}

Then, on a flat surface

\begin{equation}%22
\frac{C}{r} = 2\pi.
\end{equation}

Let us now draw a circle on the surface of a sphere, like the orange circle in Figure 8. To calculate its perimeter $C$, $\theta$ is taken to be constant in Eq. (12), which implies that $d\theta = 0$ and $\sin\theta = constant$. Integrating between $0$ and $2\pi$:

\begin{equation}%23
\int_{0}^{C} dl = r \sin\theta \int_{0}^{2\pi} d\varphi \rightarrow C=(r \sin\theta)2\pi.
\end{equation}

Let us determine the radius $\rho$ of the orange circle, which is defined as a geodesic. In this case, $\varphi = constant$ and $d\varphi = 0$ in Eq. (12). Integrating between $0$ and $\theta$:

\begin{equation}%24
\int_{0}^{\rho} dl = r\int_{0}^{\theta} d\theta \rightarrow \rho =r\theta.
\end{equation}

Taking the quotient between perimeter and radius:

\begin{equation}%25
\frac{C}{\rho} = \left( \frac{\sin\theta}{\theta}\right) 2\pi.
\end{equation}

It is easy to show (e.g., graphically) that for $\theta > 0$, $\sin\theta/\theta < 1$, so that:

\begin{equation}%26
\frac{C}{\rho} < 2\pi.
\end{equation}

A comparison of Eqs. (22) and (26) shows that the spherical surface is a non-Euclidean manifold.

\begin{figure}[h]
  \centering
    \includegraphics[width=0.3\textwidth]{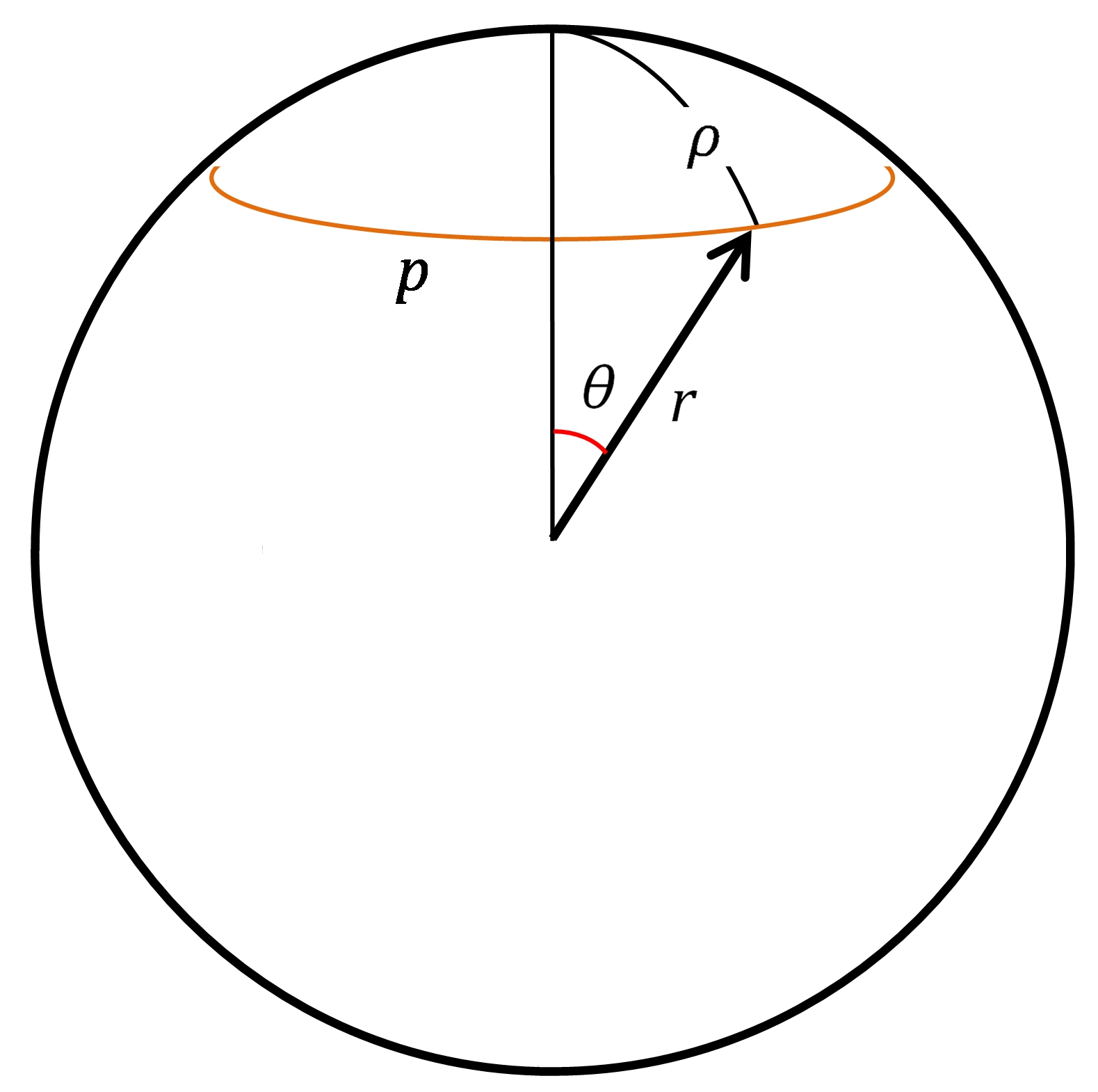}
  \caption{Circle (orange) of the perimeter $C$ and radius $\rho$ on the surface of a sphere of radius $r$.}
\end{figure}

\section{Riemannian geometry and the Minkowski metric of flat spacetime}

The theory of GR describes gravity as an effect of the curvature of spacetime, which means that without gravity the curvature is zero. This flat spacetime is the setting for the special theory of relativity (SR), which is named so because it represents a special case of GR in which there is no gravity. This spacetime is a four-dimensional manifold, that generally has three spatial and one temporal coordinate, and most of the ideas introduced above apply to it, provided we define a geodesic as a locally straight line and keep in mind that the riemannian manifolds are not exactly the same as the space-time manifolds. Therefore we must be careful in the use of the concepts, as briefly explained at the end of this section.\\ 

If the infinitesimal distance between two points in flat spacetime is denoted as $ds$, and $dl$ is used to define the infinitesimal distance between two points in flat space in coordinates $x_{1},x_{2},x_{3}$ with $x_{0} \equiv ct$ the \textit{time coordinate}, the spacetime metric can be defined as [3]:

\begin{equation}%27
ds^{2} = dx_{0}^{2} - (dx_{1}^{2} + dx_{2}^{2} + dx_{3}^{2}) = dx_{0}^{2} -dl^{2},
\end{equation}

where $c$ is the speed of light in a vacuum, and $t$ is the \textit{coordinate time}, which measures the lapse relative to a given coordinate system. If $dl$ is expressed in Cartesian coordinates, Eq. (14), where $x_{1}=x,x_{2}=y,x_{3}=z$, the \textit{Minkowski metric} can be defined as:

\begin{equation}%28
ds^{2} = c^{2}dt^{2} - (dx^{2} + dy^{2} + dz^{2}),
\end{equation}

Using Eq. (16), where $x_{1}=r,x_{2}=\theta,x_{3}=\varphi$, $dl$ can also be expressed in spherical coordinates:

\begin{equation}%29
ds^{2} = c^{2}dt^{2} - (dr^{2} + r^{2}d\theta^{2} + r^{2}\sin^{2}\theta d\varphi^{2}),
\end{equation}

In both SR and GR it is customary to use the Greek subscripts $\mu,\nu$ to define the coordinates, where $\mu,\nu = 0$ denotes the temporal coordinate, and $\mu,\nu = 1,2,3$ denotes the spatial coordinates. Then, the metric tensor associated with Eq. (28) is defined as:

\begin{equation} %30
\left[ \eta_{\mu \nu} \right] = \begin{pmatrix} \eta_{00} & \eta_{01} & \eta_{02} & \eta_{03} \\ \eta_{10} & \eta_{11} & \eta_{12} & \eta_{13} \\ \eta_{20} & \eta_{21} & \eta_{22} & \eta_{23} \\ \eta_{30} & \eta_{31} & \eta_{32} & \eta_{33} \end{pmatrix}= \begin{pmatrix} 1 & 0 & 0 & 0 \\ 0 & -1 & 0 & 0 \\0 & 0 & -1 & 0 \\ 0 & 0 & 0 & -1 \end{pmatrix}.
\end{equation}

where $\eta_{00}=1, \eta_{11}=\eta_{22}=\eta_{33}=-1$ $\mu,\nu =0$ if $\mu \neq \nu$. This is the \textit{Minkowski tensor}, which describes the flat spacetime of SR. Therefore, the Minkowski metric can be rewritten as [3]:

\begin{equation}%31
ds^{2} = \sum_{\mu=0}^{3} \sum_{\nu=0}^{3} \eta_{\mu \nu}dx_{\mu}dx_{\nu}.
\end{equation}

As in the manifolds introduced earlier, $ds$ is an invariant, but its interpretation is different. Let us consider two events that occur at the same place in a reference frame $S$, but at different times separated by an infinitesimal interval $dt$. For an observer in $S$, it must be the case that $dx=dy=dz=0$, and Eq. (28) reduces to:

\begin{equation}%32
ds=cdt \equiv cd\tau,
\end{equation}

where $\tau$ is the \textit{proper time} and can be interpreted as the lapse measured by a “wristwatch” that travels with the observer in $S$, that is, $ds$ is the “displacement in time” made by the observer on his world line. On the other hand, if Eq. (28) is used to describe a light pulse, then $dl=cdt$, so:

\begin{equation}%33
ds = 0,
\end{equation}

and the light is said to describe a \textit{null geodesic} in spacetime.\\

Before finishing this section, it is worth making a technical clarification on which we will not delve. Let us note that the tensors (17) and (30) are similar, which suggests that both describe flat manifolds. However, in Eq. (17) all the components are positive, which in mathematical terms means that it is \textit{positive definite}, while in Eq. (30) there are negative and positive components, which means that it is not positive definite. The metrics of Riemannian geometry are always positive definite, but the metrics of relativistic physics in general are not. To distinguish both cases, it is said that the space-time of relativistic physics has \textit{pseudo-Riemannian geometry}. In particular it is said that Eq. (30) is the metric tensor of a pseudo-Euclidean geometry. Taking this idea, in perfect analogy with the results presented in the previous sections it can be shown that, when the curvature is zero at every point of spacetime, it is always possible to find a coordinate system where the metric tensor takes the simple diagonal form $\left[ \eta_{\mu \nu} \right]$, Eq. (30). But, as we will see in the next chapter, when spacetime is curved, the metric tensor takes a much more complicated form that typically depends on the gravitating masses that generate the curvature.\\

To clarify and order the previous ideas it is useful to introduce the concept of \textit{signature}, which is a convention used in relativistic physics regarding the value of the signs (plus or minus) that accompany the components of the metric tensor. In the case of diagonal metrics, which will be the case we will study here, the  signature can be either $(+,-,-,-)$ or $(-,+,+,+)$, which implies that metrics are not positive definite. In Eq. (30) the signature is $(+,-,-,-)$ but it could also be $(-,+,+,+)$ without this choice altering the physical results derived from the metric. We will preferably work with the signature $(+,-,-,-)$. In Riemannian geometry the concept of signature is irrelevant, since all the components of a metric tensor are positive, which implies that Riemannian metrics are always positive definite.

\section{Final comments}

Since Galileo Galilei laid the foundations for the scientific method, physics and mathematics have maintained a close and fruitful relationship, which has contributed to their mutual growth, allowing physics to provide us with an increasingly detailed and accurate image of reality. One of the most emblematic examples is Einstein's use of Riemannian geometry to describe gravity as a geometric effect of the curvature of spacetime. It was a close friend of Einstein, the mathematician Marcel Grossman, who introduced Einstein to the study of Riemannian geometry [5].\\

More than a century after the brilliant synthesis between relativity and Riemannian geometry carried out by Einstein, the fruits are visible: black holes, neutron stars, gravitational lenses, active galaxies, gravitational waves, etc. How Einstein consummated the marriage between relativity and Riemannian geometry will be discussed in greater detail in the second part of this work.

\section*{Acknowledgments}
I would like to thank to Daniela Balieiro and Michael Van Sint Jan for their valuable comments in the writing of this paper. I would also like to thank the referees, whose comments and suggestions have allowed me to significantly improve this paper.

\section*{References}

[1] K. S. Thorne, Black Holes and Time Warps: Einstein’s Outrageous Legacy, WW Norton and Co, New York, 2014.

\vspace{2mm}

[2]	J. Pinochet, Classical Tests of General Relativity Part I: Looking to the Past to Understand the Present, Physics Education. 55 (2020) 65016.

\vspace{2mm}

[3]	R. Lambourne, Relativity, Gravitation and Cosmology, Cambridge University Press, New York, 2010.

\vspace{2mm}

[4]	B. Schutz, A First Course in General Relativity, 2nd ed., Cambridge University Press, New York, 2009.

\vspace{2mm}

[5]	A. Einstein, Die Grundlage der allgemeinen Relativitätstheorie, Annalen Der Physik. 354 (1916) 769–822.

\vspace{2mm}

\end{document}